\documentclass[12pt,preprint]{aastex}

\shorttitle{Large-Scale Coronal Structure}
\shortauthors{Ingleby et al}

\begin{document}

\title{Probing the Large Scale Plasma Structure of the Solar Corona with Faraday Rotation Measurements}

\author{Laura D. Ingleby, Steven R. Spangler, and Catherine A. Whiting}
\affil{Department of Physics and Astronomy, University of Iowa, Iowa City, IA 52242}

\begin{abstract}
Faraday rotation measurements of the solar corona made with the Very Large Array (VLA) at frequencies of 1465 and 1665 MHz are reported.  The measurements were made along 20 lines of sight to 19 extragalactic radio sources in March and April, 2005. The closest heliocentric distances of the lines of sight ranged from 9.7 to 5.6 $R_{\odot}$. Measured rotation measures range from -25 to +61 rad/m$^2$.  The purpose of these observations is to probe the three dimensional structure of the coronal plasma in the heliocentric distance range $5-10 R_{\odot}$, and particularly the strength and structure of the coronal magnetic field.  The measured rotation measures are compared with two types of models for the coronal plasma structure.  The first model utilizes an algebraic expression for the rotation measure, based on power law expressions for the plasma density and magnetic field as a function of heliocentric distance.  The only input parameter which depends on current solar conditions is the position of magnetic neutral line in the corona.  The second set of models employed independent measurements of coronal density and magnetic field along each line of sight.  These independent data sets are estimates of the coronal density at $r=3.0 R_{\odot}$ obtained from the SOHO/LASCO C2 coronagraph,  and potential field estimates of the coronal magnetic field at a heliocentric distance of $3.25 R_{\odot}$.  These estimates of the density and magnetic field are extrapolated to the larger heliocentric distances probed by the radio observations with power law functions of the radial distance.  For the majority of the lines of sight,  the observed rotation measures are reasonably well represented by the predictions from either type of model.  However, 4 of the 20 lines of sight have large observed-model residuals.  These anomalous sources do not seem to be affected by coronal mass ejections (which did occur during the program),  but there may be an association between anomalous rotation measures and the presence of streamers along the line of sight.  Our primary result on the magnitude of the coronal magnetic field is taken from the second set of models, which incorporate more information on the coronal state at the time of the observations. The magnitude of the  field necessary to reproduce the majority of the observations is in the range 30-78 milliGauss at a fiducial distance of $6.2 R_{\odot}$ (46-120 milliGauss at $5 R_{\odot}$), with a smaller, preferred range of 30 - 34 milliGauss at $6.2 R_{\odot}$ (46-52 mG at $5 R_{\odot}$).
\end{abstract}



\section{Introduction}
\begin{figure}
\epsscale{.75}
\plotone{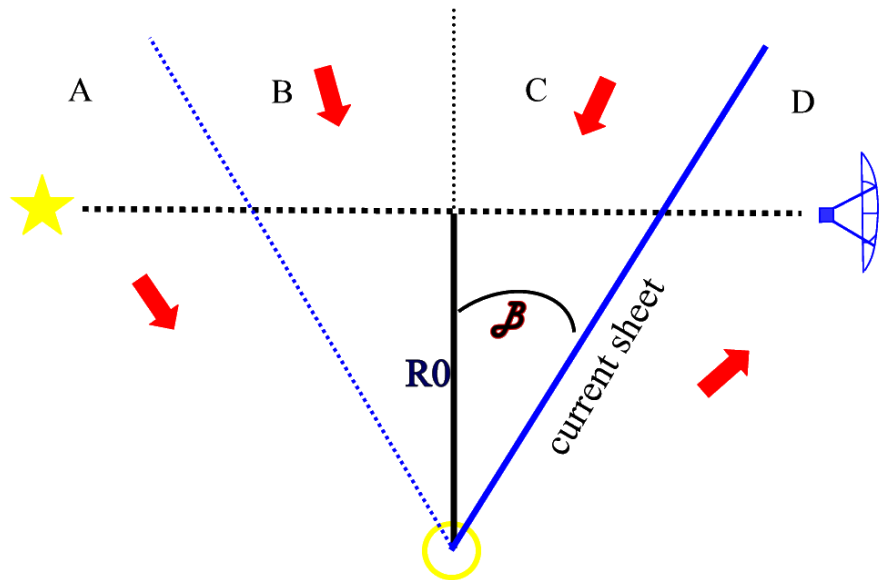}
\caption{The geometry of a coronal Faraday rotation measurement. The line of sight extends from a radio source, through the corona, to a receiving station.  The line of sight passes at a closest distance $R_0$ which is referred to as the impact parameter.  The figure illustrates an idealization which will be employed in this paper, in which the coronal magnetic field is radial.  The angle $\beta$ is a coordinate which defines the position along the line of sight. The value of the rotation measure depends sensitively on the location of the magnetic neutral line along the line of sight (given by the angle $\beta_c$).} 
\label{cartoon1}
\end{figure}

One of the few observational techiques for measurement of the coronal magnetic field is Faraday rotation of a trans-coronal, linearly polarized source of radio waves.  This technique has been used extensively in the past, and  illustrative references  are \citet{Bird90,Patzold87,Sakurai94a,Mancuso99,Mancuso00} and \cite{Spangler05}. The geometry of the situation is shown in Figure~\ref{cartoon1}.

 The change in position angle $\Delta \chi$ which defines Faraday rotation is given by 
\begin{equation}
\Delta \chi = \left[ \left( \frac{e^3}{2 \pi m_e^2 c^4}\right) 
              \int_L n_e \vec{B} \cdot \vec{dz} \right] \lambda^2
\end{equation}
The variables in equation (1) are as follows.  The fundamental physical constants of $e, m_e, \mbox{ and } c$ are, respectively, the fundamental charge, the mass of an electron, and the speed of light.  The electron density in the plasma is $n_e$, and $\vec{B}$ is the vector magnetic field.  The incremental vector $\vec{dz}$ is a spatial increment along the line of sight, which is the path on which the radio waves propagate. Positive $z$ is in the direction from the source to the observer. The subscript $L$ on the integral indicates an integral along the line of sight.  Finally, $\lambda$ indicates the wavelength of observation.  The term in square brackets is called the rotation measure, and is the physical quantity retrieved in Faraday rotation measurements. 
One of the goals of the works cited above has been to determine the magnitude of the coronal field as a function of heliocentric distance.   A more complete understanding of the coronal field includes determining its global structure, i.e. the dependence on heliographic coordinates $(r,\theta,\phi)$.  

As may be seen in equation (1), deducing the coronal field from Faraday rotation measurements is complicated by two factors.  The first is that the magnetic field is contained within a path integral along the line of sight from the source of radio waves to the observer,  and that this integrand can be positive as well as negative at different points along the line of sight.  It is therefore completely possible that radio waves can traverse a region of strong magnetic field, and yet have zero Faraday rotation if the integrand has equal positive and negative parts.  

The second complication is that the integrand depends not only on the coronal magnetic field, but is also dependent on the plasma density.  To extract useful information on the coronal field and circumvent these difficulties, it is necessary to have independent information on the plasma density in the corona and along the line of sight,  and also have independent information on the geometry of the magnetic field,  consisting of (at least) knowledge of where the field is of positive and negative polarity. This analysis is made possible by the fact that there exists a substantial body of independent information on the plasma density in the solar corona \citep[see][]{Bird94,Guhathakurta96}, including information on the coronal density distribution at the time of our observations.  

The goal of this paper is to infer the coronal magnetic field from a large set of Faraday rotation measurements,  in which the lines of sight probe different parts of the corona, i.e. regions substantially separated in heliographic latitude and longitude.  

\begin{figure}
\includegraphics[angle=90,width=35pc]{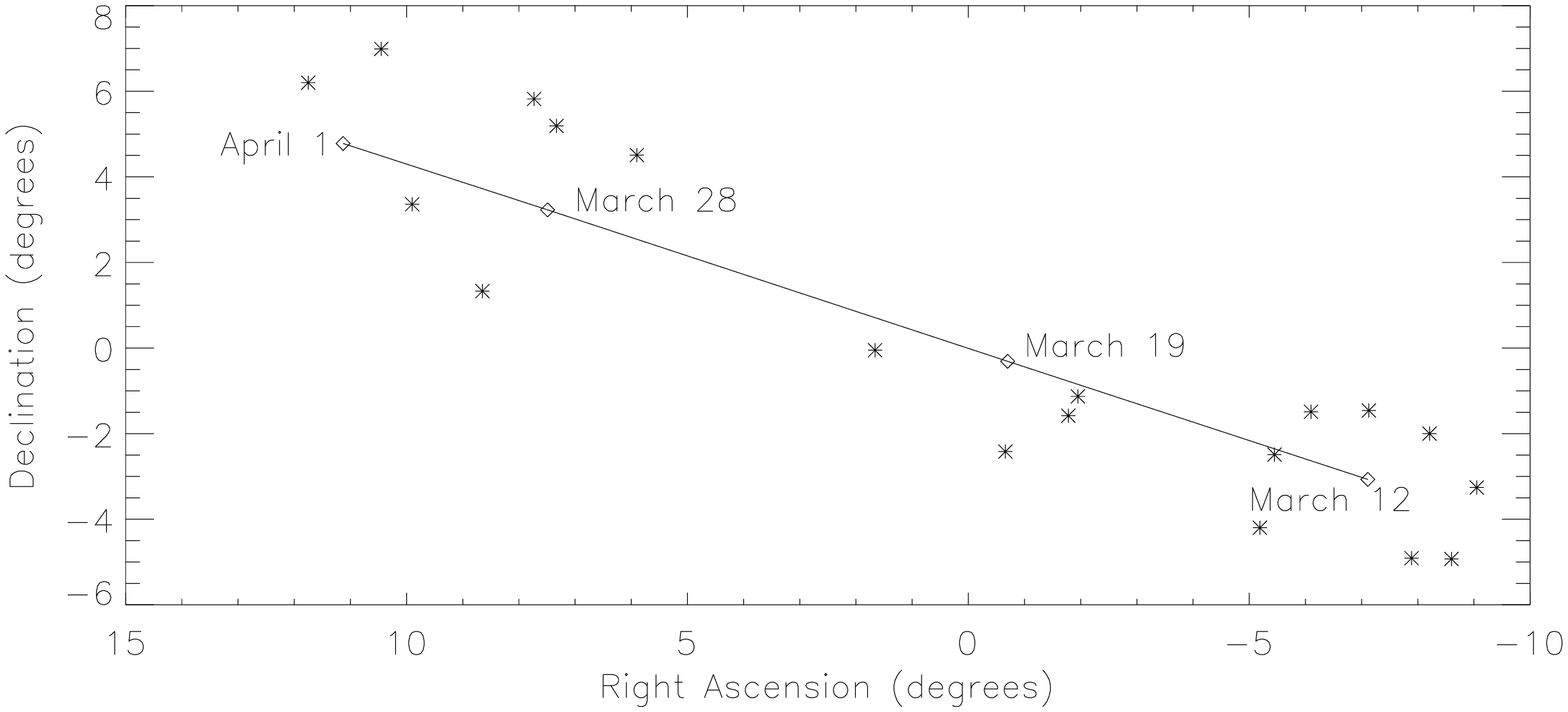}
\caption{Location of the Sun on the four days of observation in March and April, 2005 (diamond symbols).  Locations of the sources are indicated by asterisks.  Each source was observed on a single day (easily inferred from the figure)  except for the source 0039+033, which was observed on both March 28 and April 1, 2005.  A list of the sources observed, and the date of observation for each source, is given in Table 1. }
\label{constellation1}
\end{figure}

 We present measurements along 20 lines of sight through the corona during a period which corresponded to approximately one solar rotation. A novel feature of this investigation is our utilization of independent information on the coronal plasma density, as well as the geometry and structure of the coronal magnetic field at the epoch of observation. The lines of sight probed, relative to the Sun on the days of observation, are shown in Figure~\ref{constellation1}.  

The scope of this investigation is similar to that of \cite{Mancuso00}.  The difference is that the present project includes more lines of sight (20 versus 13) and was undertaken at a different time. The investigation of \cite{Mancuso00} was done at the very beginning of Cycle 23, while the present project was done near the end of that cycle.  Furthermore, the form of modeling of the expected rotation measures in the present investigation is different, and makes fuller use of independent data on the state of the corona.   

\section{Observations} All observations were made with the Very Large Array radio telescope of the National Radio Astronomy Observatory\footnote{The Very Large Array is an instrument of the National Radio Astronomy Observatory. The NRAO is a facility of the National Science Foundation, operated under cooperative agreement with Associated Universities, Inc.}. 
\subsection{Radio ``Constellations'' and the Sun in March and April, 2005}
The dates for the observations were chosen to maximize the number and distribution of lines of sight through the corona,  given practical limitations on observing time.  This  was achieved in two ways.  
\begin{enumerate}
\item The observations were made in four sessions,  spaced approximately one week, or a quarter of a solar rotation, apart.  As a result, the lines of sight probed sets of heliographic longitudes spaced by roughly 90 degrees.  
\item The dates for the sessions were chosen when the Sun was located in a ``constellation'' of radio sources suitable for polarization measurements.  These constellations were comprised of sources chosen from the NRAO NVSS \citep{Condon98} survey on the basis of polarized flux density and proximity to the Sun. We required program sources to have at least 10 mJy of polarized flux at 1465 MHz (L band), and be sufficiently compact for observations with the VLA B array. Proximity to the Sun was determined from the dual requirements that the source be close enough for detectable coronal Faraday rotation at L band, but far enough that solar contributions to the system temperature were acceptable. On the basis of prior experience with coronal Faraday rotation, this meant sources with solar elongations from approximately 1.25 - 2.5 degrees.  
\end{enumerate}
The days which were chosen were March 12, March 19, March 28, and April 1, 2005.  For each session, observations were made from approximately 15h - 23h UT.  

The locations of the sources relative to the Sun are shown in Figure~\ref{constellation1}.  The roster of sources is given in Table 1.  Columns 1 and 2 give the source name and the date of observation.  Columns 3 and 4 give the Right Ascension and Declination of each source (J2000 coordinates), and Columns 5 and 6 give the solar elongation (angle between the source and the center of the Sun) and position angle of the source at 20h UT on the date of observation. Note that the source 0039+033 is listed twice, as it was favorable for observation on both March 28 and April 01. 
\begin{deluxetable}{crrrrr}
\tabletypesize{\scriptsize}
\tablecaption{Source Roster\label{tbl-1}}
\tablewidth{0pt}
\tablehead{\colhead{Source} & \colhead{Date} &  \colhead{RA(J2000)} & \colhead{DEC(J2000)} & \colhead{$R_0 (\circ)$}  & \colhead{$\chi (\circ)$} }
\startdata
2323-033 & Mar12 & $23h23m31.98s$ & $-03^{\circ}17^{'}05.4^{''}$ & 1.95 & 264.5\\
2325-049 & Mar12 & $23h25m19.58s$ & $-04^{\circ}57^{'}36.9^{''}$ & 2.39 & 218.7 \\
2326-020 & Mar12 & $23h26m54.47s$ & $-02^{\circ}02^{'}10.3^{''}$ & 1.53 & 314.1 \\
2328-049 & Mar12 & $23h28m11.99s$ & $-04^{\circ}56^{'}08.6^{''}$ & 1.99 & 202.8 \\
2331-015 & Mar12 & $23h31m13.44s$ & $-01^{\circ}29^{'}13.5^{''}$ & 1.61 & 359.3 \\
2335-015 & Mar12 & $23h35m20.43s$ & $-01^{\circ}31^{'}10.1^{''}$ & 1.87 &  32.6 \\
2337-025 & Mar12 & $23h37m57.36s$ & $-02^{\circ}30^{'}58.4^{''}$ & 1.76 & 70.7 \\
2338-042 & Mar12 & $23h38m59.19s$ & $-04^{\circ}13^{'}33.8^{''}$ & 2.23 & 120.5 \\
2351-012 & Mar19 & $23h51m56.21s$ & $-01^{\circ}09^{'}16.3^{''}$ & 1.49 & 236.8 \\
2352-016 & Mar19 & $23h52m49.85s$ & $-01^{\circ}36^{'}13.2^{''}$ & 1.63 & 219.1 \\
2357-024 & Mar19 & $23h57m05.71s$ & $-02^{\circ}26^{'}36.0^{''}$ & 2.11 & 179.0 \\
0006-001 & Mar19 & $00h06m22.60s$ & $-00^{\circ}04^{'}25.1^{''}$ & 2.40 & 83.7 \\
0023+045 & Mar28 & $00h23m19.74s$ & $+04^{\circ}28^{'}48.6^{''}$ & 2.04 & 308.8 \\
0029+052 & Mar28 & $00h29m03.59s$ & $+05^{\circ}09^{'}34.1^{''}$ & 1.96 & 355.3 \\
0030+058 & Mar28 & $00h30m40.81s$ & $+05^{\circ}47^{'}14.9^{''}$ & 2.60 & 5.4 \\
0034+013 & Mar28 & $00h34m19.70s$ & $+01^{\circ}18^{'}07.9^{''}$ & 2.22 & 148.7 \\
0039+033 & Mar28 & $00h39m20.35s$ & $+03^{\circ}19^{'}49.2^{''}$ & 2.41 & 87.0 \\
0039+033 & Apr01 & $00h39m20.35s$ & $+03^{\circ}19^{'}49.2^{''}$ & 1.88 & 220.8 \\
0041+070 & Apr01 & $00h41m32.37s$ & $+06^{\circ}57^{'}50.3^{''}$ & 2.31 & 342.9 \\
0046+067 & Apr01 & $00h46m44.01s$ & $+06^{\circ}10^{'}08.4^{''}$ & 1.54 & 23.5 \\
\enddata
\end{deluxetable}

\subsection{Acquisition and Analysis of VLA Data}
The methods of observation and data reduction were very similar to those previously used and reported in \cite{Sakurai94a,Sakurai94b,Mancuso99,Mancuso00} and \cite{Spangler05}. As previously noted, the present project is more extensive that the prior investigations, and had four sessions in which sources were observed through the corona.  In addition, a fifth session was scheduled on May 29, 2005.  This was a reference session, in which the polarization of the sources in the absence of the corona was measured.   The following enumerates some features of the observations which were novel, or which are particularly worth emphasizing here. Further elaboration and justification of some of these points may be found in the aforementioned papers.   
\begin{enumerate}
\item Observations were made simultaneously at frequencies of 1465 (20cm) and 1665 (18cm) MHz, with intermediate frequency bandwidths of 50 and 25 MHz, respectively.
\item Observations were made when the array was in the ``B array'',  which has a synthesized beam of approximately 4-5 arcseconds, FWHM. 
\item Program sources were observed for scans of approximately 6 minutes duration.  Depending on the session (i.e. the number of sources in the constellation for a particular day), the number of scans of a source ranged from 6 to 17. 
\item Ionospheric Faraday rotation was estimated and corrected for via the AIPS task TECOR, which makes use of an ionospheric model, appropriate to the time of the observations, obtained from GPS data.  
\item Determination of the instrumental polarization (contained in the D factors for each antenna) was done from observations of the primary phase calibrator.  However, in each session a spare calibrator was also observed and used to independently solve for the D factors.  For each session, the agreement between D factors obtained from the two sets of data was excellent, with a level of agreement between the independent estimates of the D factors similar to that illustrated in Figure 2 of \cite{Sakurai94b}.  This insured fidelity of the polarization calibration. The primary and secondary polarization calibrators were 2330+110 and 0059+001 for March 12 and March 19,  0059+001 and 2330+110 on March 28, and 0022+002 and 2330+110 on April 1.   
\item The source 3C138 was observed for calibration of the polarization position angle. In each observing session, 3C138 was observed twice, with the  scans separated by several hours.  This was done to insure reliability of the A-C  phase difference of the array, which sets the origin of the polarization position angle. 
\item In addition to polarization calibration observations, in each session two observations were made of an extended radio source far from the Sun.  One scan was made before transit and the other after.  Both scans were mapped and the distribution of polarization position angle over the source measured. These observations were made to insure that there was no residual, uncalibrated systematic error in the polarimeter characteristics or the ionosphere Faraday rotation, which could corrupt the time dependence and mean values of the coronal rotation measures.  In no case was there a systematic difference of more than a few degrees between the two position angle maps of the source. 
\item The mapping procedure consisted of one iteration of phase-only self calibration.  
\end{enumerate}
The reference observations to determine the intrinsic polarization position angle maps\footnote{These reference maps of the polarization position angles are determined by the intrinsic polarized emission mechanism, Faraday rotation within the sources and surrounding plasma media, and the interstellar medium of the Milky Way.} were made on May 29, 2005, when the Sun was far removed from all of the program sources.  Polarimeter calibration was carried out as normal, and each of the program sources was observed. 
For each of the 19 program sources, maps of the Stokes parameters I, Q, and U at both 1465 and 1665 MHz were made. These were used to generate reference maps of the derived polarization parameters, the polarization position angle $\chi \equiv \frac{1}{2}\tan^{-1} \left( \frac{U}{Q}\right)$ and the polarized intensity $L \equiv \sqrt ( Q^2 + U^2 )$.  

During the ``program sessions'' when sources were being observed through the corona,  a full set of maps was made for each scan on every source, and visually examined.  The purpose of this exercise was to check for evidence of significant time-dependent Faraday rotation during the eight hour observing session.  Although this program was not intended to study time-variable Faraday rotation \citep{Sakurai94a,Mancuso99,Spangler05},  averaging data for a source with strongly-varying Faraday rotation could bias the derived polarization parameters.  In especially pronounced cases, such as that of the source 3C228 on August 16, 2003 \citep{Spangler05}, excessive averaging can result in apparent depolarization.

Two of the 20 lines of sight displayed substantial variation in the Faraday rotation over the 8 hour duration of the observing session.  These were 2351-012 on March 19 and 0046+067 on April 1.  Further discussion of these observations is given in Section 3.1.  For the remaining 18 lines of sight,  a visual examination of the full set of polarization position angle maps showed little or no discernible Faraday rotation during the session.  In these cases, all of the data in the session were used to make a single set of maps which thus possessed superior sensitivity and fidelity resulting from the much better $(u,v)$ plane coverage.  

\begin{figure}[h]
\epsscale{0.60}
\includegraphics[angle=-90,width=12pc]{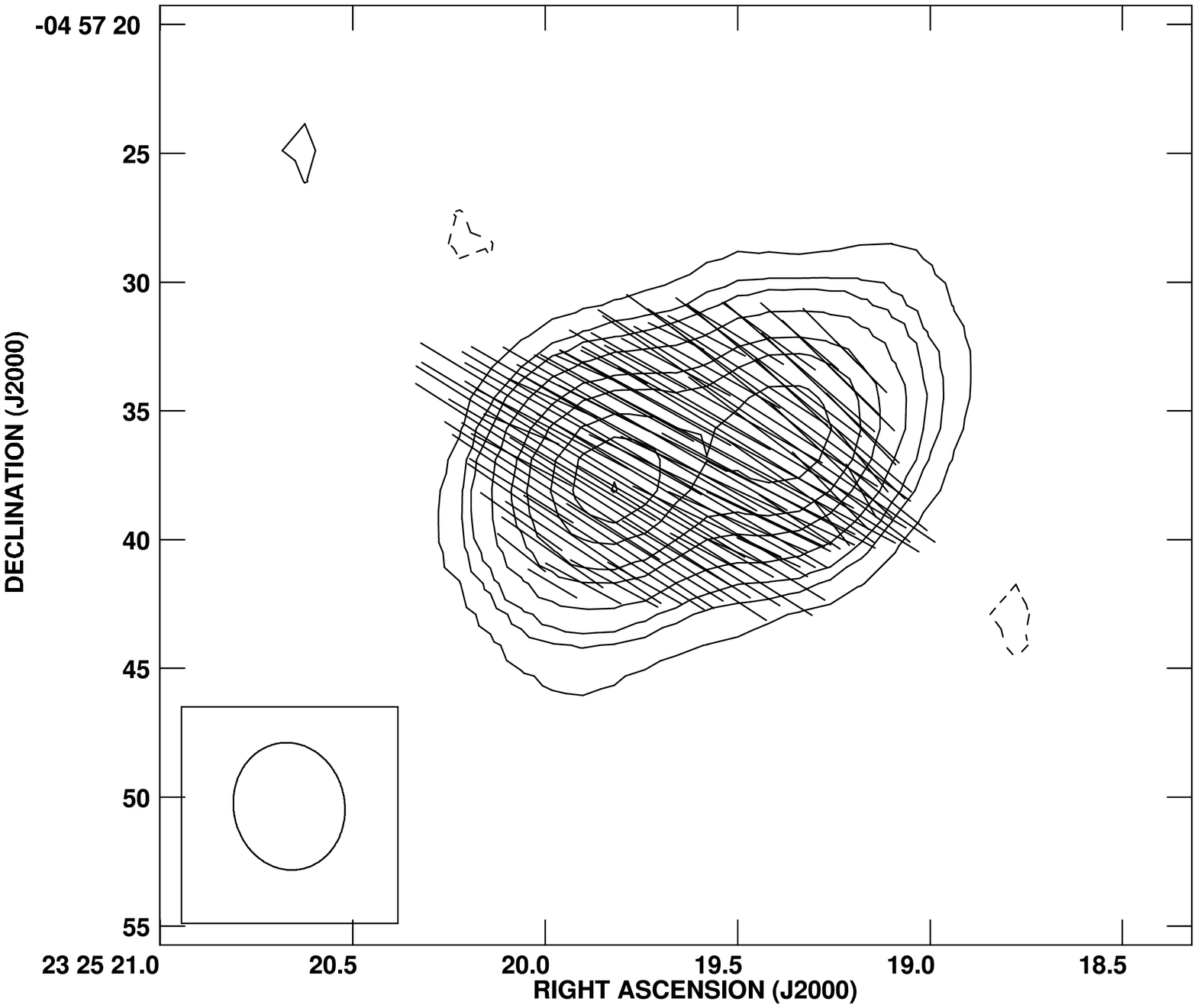}
\includegraphics[angle=-90,width=12pc]{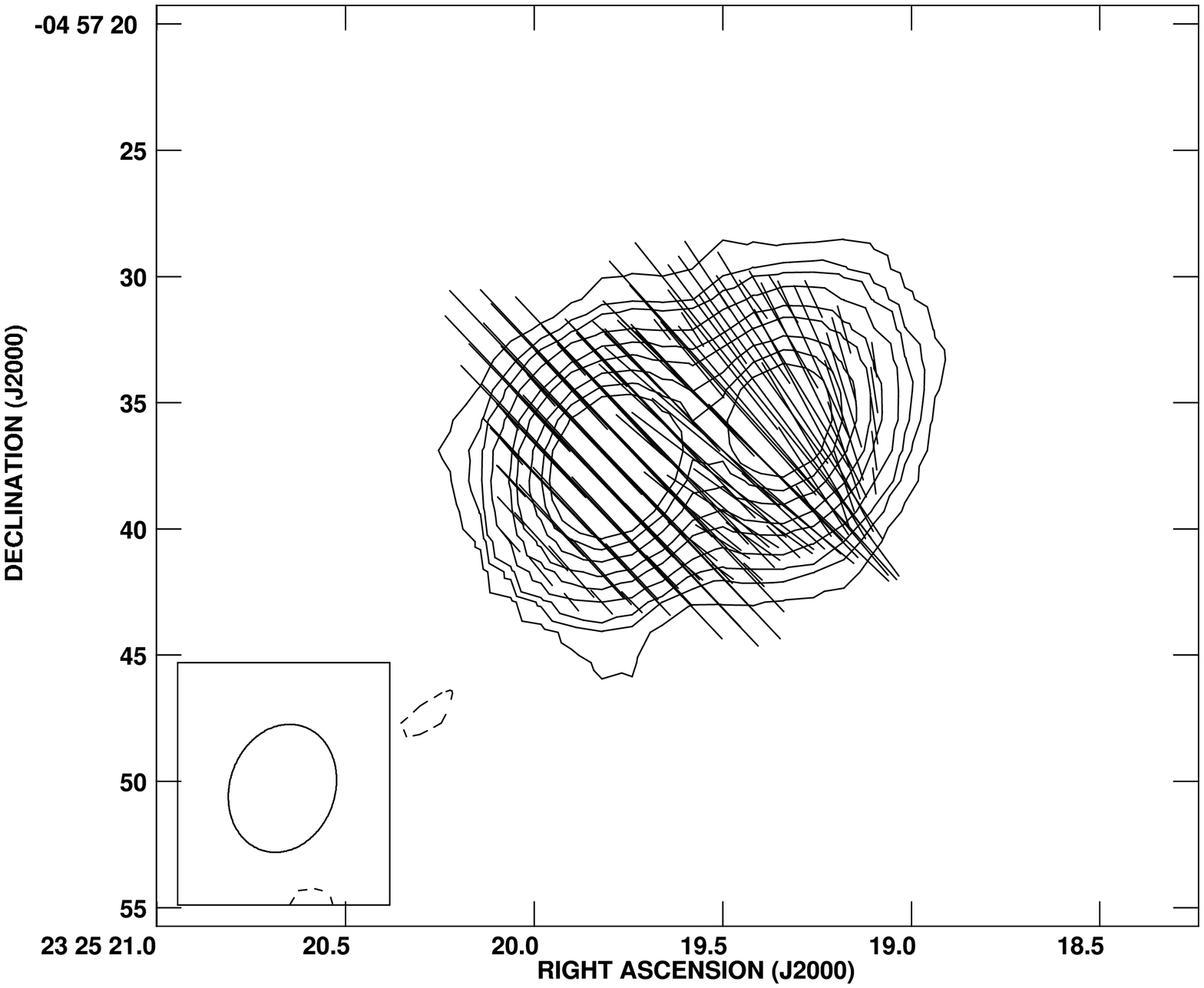}
\includegraphics[angle=-90,width=12pc]{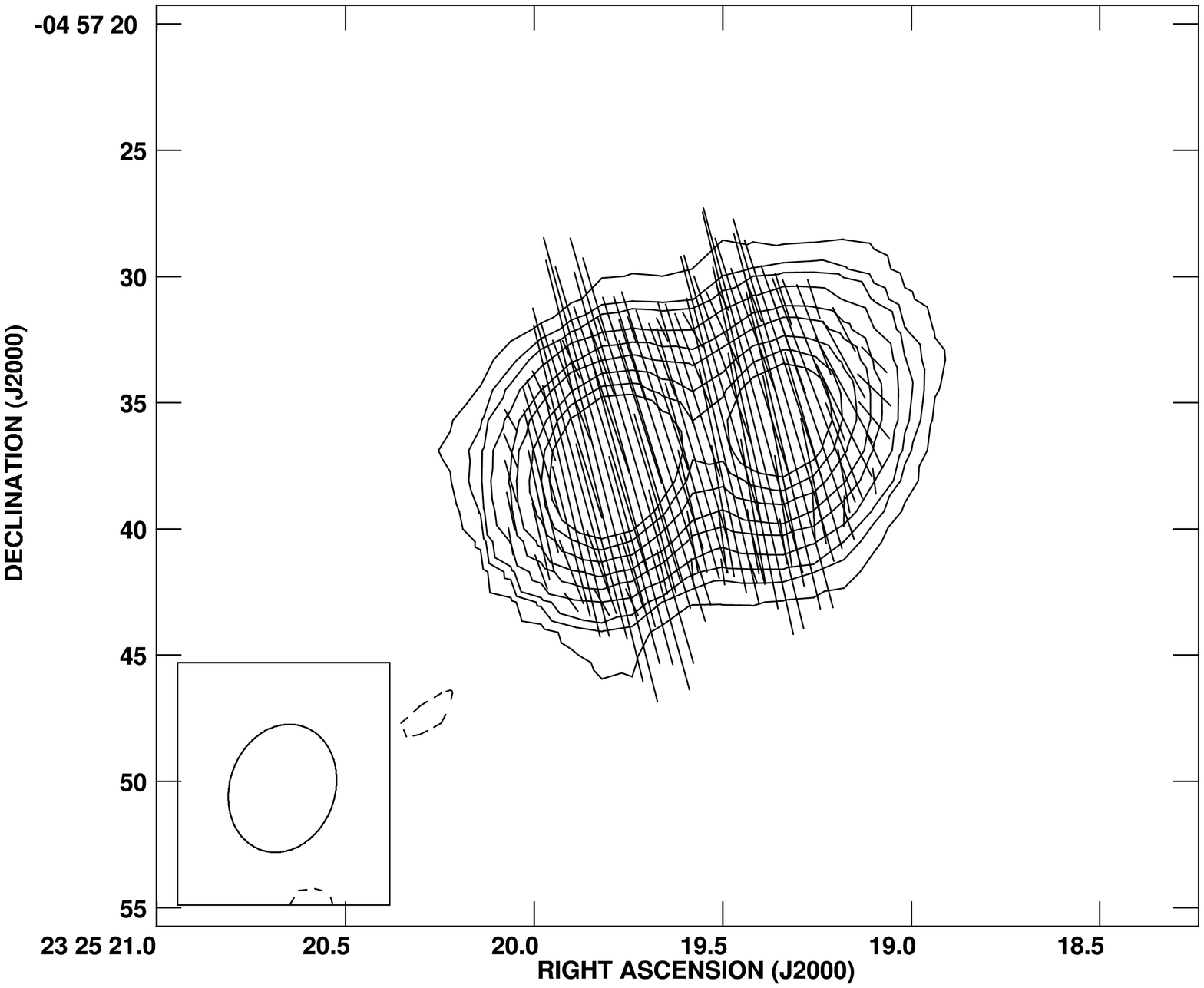}
\caption{Illustration of data used to obtain the coronal Faraday rotation for one of the lines of sight, that to 2325-049 on March 12. In all three panels,  the lines correspond to the polarization position angle at that part of the source,  and the contours are of total intensity (Stokes Parameter I).  The left panel shows the 1465 MHz polarization position angle map on March 12, when the line of sight passed through the corona.  The middle panel shows a similar map of the source on May 29, when the corona was far from the source. The right panel shows the position angle difference between the two maps, $\Delta \chi$, which is the Faraday rotation due to the corona. For both components of the source, the Faraday rotation is about $14^{\circ}$, corresponding to a rotation measure of about 6 radians/m$^2$. The resolution of all three maps is about 5 arcseconds (angular diameter FWHM of the restoring beam). }
\label{2325DXIF1}
\end{figure}

An illustration of one of our coronal Faraday rotation measurements is shown in Figure~\ref{2325DXIF1}.  The three panels show the polarization position angle map at 1465 MHz when the source was observed through the corona on March 12 (left panel), during the reference period on May 29 (middle panel) and the position angle difference $\Delta \chi$ map derived from the two maps (right panel).  There is clearly a rotation for both source components corresponding to $14^{\circ}$ of position angle, or a rotation measure of about 6 rad/m$^2$ (see Table 2 for precise values).   

\section{Observational Results}
For each of the sources, $\Delta \chi$ measurements were made as described above for as many source components as permitted by the signal-to-noise ratio. For example, in Figure 3  there are two such source components. The data for the different components were examined for {\em differential Faraday rotation}, in which there is a difference in the rotation measure to two source components separated by angles from a few arcseconds to tens of arcseconds \citep{Spangler05}.  No convincing case of differential Faraday rotation was seen, and in fact was not expected in view of the averaging of the data over the duration of the session.  The data were also checked for adherence to the $\lambda^2$ dependence expected for Faraday rotation; the 1665 MHz $\Delta \chi$ values were required to be 29 \% less than the 1465 MHz values.  With the exception of a small handful of outlyers, the data adhered to the expected relationship within the errors. 
For each source, we obtained a mean rotation measure by averaging the RM values for each source component and the two frequencies.  For sources possessing at least 4 such measurements (two components at two frequencies), we calculated a second estimate of the mean by excluding the most discrepant measurement. The rotation measures reported here and used in the subsequent analysis employed this second estimate, when available, although the means incorporating all measurements are only slightly different.  

The results of our measurements are shown in Table 2, which provides the primary data set for this paper. 

\begin{deluxetable}{crrr}
\tabletypesize{\scriptsize}
\tablecaption{Rotation Measure Measurements\label{tbl-2}}
\tablewidth{0pt}
\tablehead{\colhead{Source} & \colhead{Date} & \colhead{No. Comp} & \colhead{$RM$ (rad/m$^2$)} }
\startdata
2323-033 & Mar12 & 1 & $61.1 \pm 1.0$ \\
2325-049 & Mar12 & 2 & $6.0 \pm 0.7$\\
2326-020 & Mar12 & 3 & $-2.1  \pm 1.0$\\
2328-049 & Mar12 & 2 & $-3.7 \pm 0.5$\\
2331-015 & Mar12 & 2 & $-7.4 \pm 1.6$\\
2335-015 & Mar12 & 1 & $-13.7 \pm 0.3$\\
2337-025 & Mar12 & 1 & $-12.5  \pm 0.5$\\
2338-042 & Mar12 & 2 & $-5.2 \pm 0.2$\\
2351-012 & Mar19 & 1 & $-27.4 \rightarrow -4.6$\\
2352-016 & Mar19 & 2 & $-24.5 \pm 0.6$\\
2357-024 & Mar19 & 1 & $3.1 \pm 0.2$\\
0006-001 & Mar19 & 2 & $2.7 \pm 0.2$\\
0023+045 & Mar28 & 1 & $5.2 \pm 0.1$\\
0029+052 & Mar28 & 1 & $-2.5 \pm 0.2$\\
0030+058 & Mar28 & 2 & $0.2 \pm 0.4$\\
0034+013 & Mar28 & 2 & $1.6 \pm 0.6$\\
0039+033 & Mar28 & 2 & $2.6 \pm 0.6$\\
0039+033 & Apr01 & 2 & $-1.4 \pm 0.8$\\
0041+070 & Apr01 & 2 & $-0.2 \pm 0.6$\\
0046+067 & Apr01 & 3 & $-1.9 \rightarrow -13.3$\\
\enddata
\tablecomments{Sources 2351-012 on March 19, and 0046+067 on April 1 showed Faraday 
               Rotation variability during the observing session. }
\end{deluxetable}
Columns 1 and 2 give the source name and the date of the observation.  Column 3 gives the number of structural components in the source which provided rotation measure values, and Column 4 gives the mean rotation measure value on that line of sight.  In the cases of the sources 2351-012 on March 19, and 0046+067 on April 1, the rotation measure was found to vary significantly during the observing session,  and the entry in Column 4 gives the range of values observed.

\subsection{Sources with Time-Variable Rotation Measure}
The project described in this paper was intended to probe the overall, quasi-static structure of the corona rather than smaller scale structure and turbulence.  As such, we were interested in measuring the average rotation measure along a line of sight rather than temporal variability, as was the goal in previous studies of ours, such as \cite{Sakurai94a,Mancuso99} and \cite{Spangler05}.  Nonetheless, as mentioned in Section 2.2, source polarization maps were examined on a scan-by-scan basis to identify cases in which the rotation measure changed substantially during the eight hour duration of the observing session. 

Two such sources were noted, 2351-012 on March 19, in which the rotation measure changed by 23 radians/m$^2$ during the session, and 0046+067 on April 1, in which the rotation measure changed by 11 radians/m$^2$.  Changes of this order over a time scale of several hours are common, and have been reported in the studies referenced above.  We have a high degree of confidence in the reality of these variations, since other program sources on both of these days, observed under the same conditions and with the same polarimeter calibration, showed no such variation.  

A consideration of possible coronal sources of this variability is given in Section 4.3.  
\subsection{Mean Rotation Measure Measurements}
\begin{figure}
\includegraphics[angle=90,width=40pc]{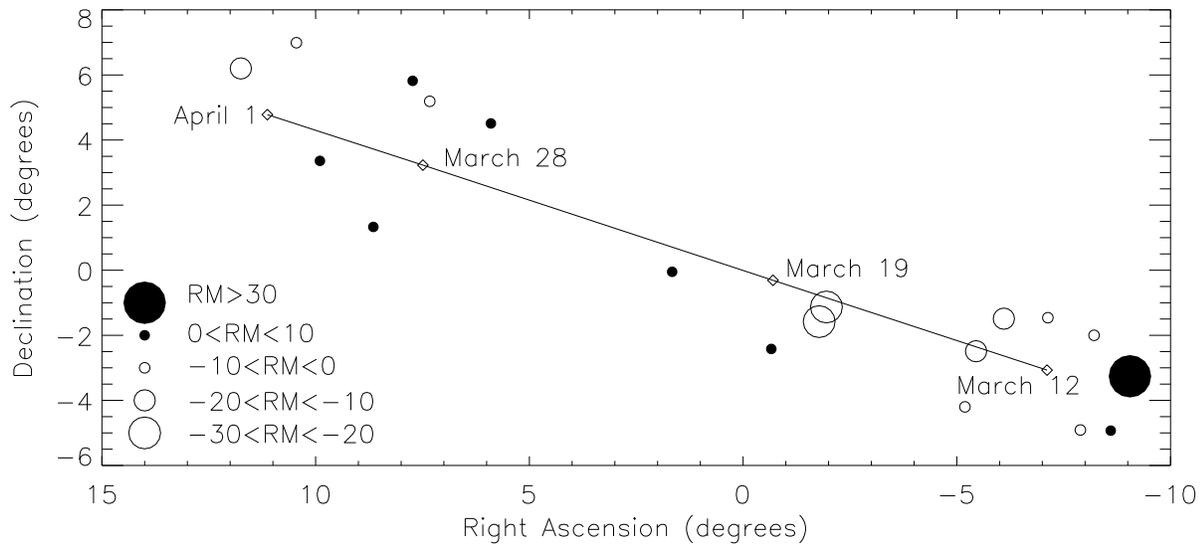}
\caption{Representation of measured rotation measures on a sky chart.  The format is the same as that of Figure~\ref{constellation1}, except the symbols now represent the rotation measure observed for each source.  The size of the plotted circle corresponds to the absolute magnitude of the rotation measure, as defined in the legend.  Solid circles correspond to positive rotation measures, and open circles represent negative rotation measures. }
\label{constellation2}
\end{figure}

For 18 of the 20 lines of sight, the rotation measure did not vary during the observing session at the level reported for 2351-012 and 0046+067.  The rotation measures for these sources, as well as mean values for the two RM-variable sources, are displayed in map format in Figure~\ref{constellation2}.  
The rotation measures for our sample, tabulated in Table 2 and plotted in Figure~\ref{constellation2}, range from -27.4 to +61.1 radians/m$^2$, although most measurements yielded a value with a much smaller absolute magnitude.  Figure~\ref{constellation2} illustrates that substantial differences in magnitude and even sign of the rotation measure can occur even for closely-spaced lines of sight.  
\section{Comparison of Observations with Model Coronae}
The goal of this project is to determine plasma physical parameters in the solar corona, particularly the strength and structure of the coronal magnetic field, from analysis of our set of Faraday rotation measurements. We do this in this section by calculating model rotation measures and comparing them with the observations. The models consist of simplified descriptions of the plasma density and magnetic field in the corona. The model rotation measures are then calculated along the exact lines of sight probed.  

Two types of models are considered in the present paper.  Section 4.1 discusses simple algebraic models which nonetheless incorporate some information about the Sun at the time of our observations.  The second class of models, discussed in Section 4.2, utilize synoptic, solar-rotation-averaged data from the time of our observations (Carrington rotations 2027 and 2028) to provide an estimate of the expected rotation measure.   
\subsection{Analytic Coronal Models}
As discussed in \cite{Patzold87} and \cite{Patzold98}, if one assumes a radial coronal magnetic field, and single power law expressions for the radial dependence of plasma density and magnetic field strength, a simple analytic expression can be obtained for the expected rotation measure. This idea was elaborated and applied to similar Faraday rotation measurements of the source 3C228 in August, 2003 \citep[Spangler et al 2007, in preparation]{Spangler05}.  Assume that the electron density and magnetic field in the corona are given by 
\begin{eqnarray}
n_e(r) = N_0 \left( \frac{r}{R_{\odot}} \right)^{-\alpha}  \\
\vec{B}(r) = B_0 \left( \frac{r}{R_{\odot}} \right)^{-\delta} \hat{e}_r  
\end{eqnarray}
These expressions are used in eq (1), and the integration is performed along the line of sight as illustrated in Figure~\ref{cartoon1}. Given equations (2) and (3), a nonzero rotation measure requires that $B$ reverse itself at some point along the line of sight.  This reversal is characterized by the angle $\beta_c$ along the line of sight, as illustrated in Figure~\ref{cartoon1}.  The parameter $\beta_c$ will differ for each line of sight, and will depend (in addition to the details of the specific line of sight) on the structure of the coronal field at the time of the observations.  

\begin{figure}
\epsscale{0.70}
\includegraphics[angle=-90,width=28pc]{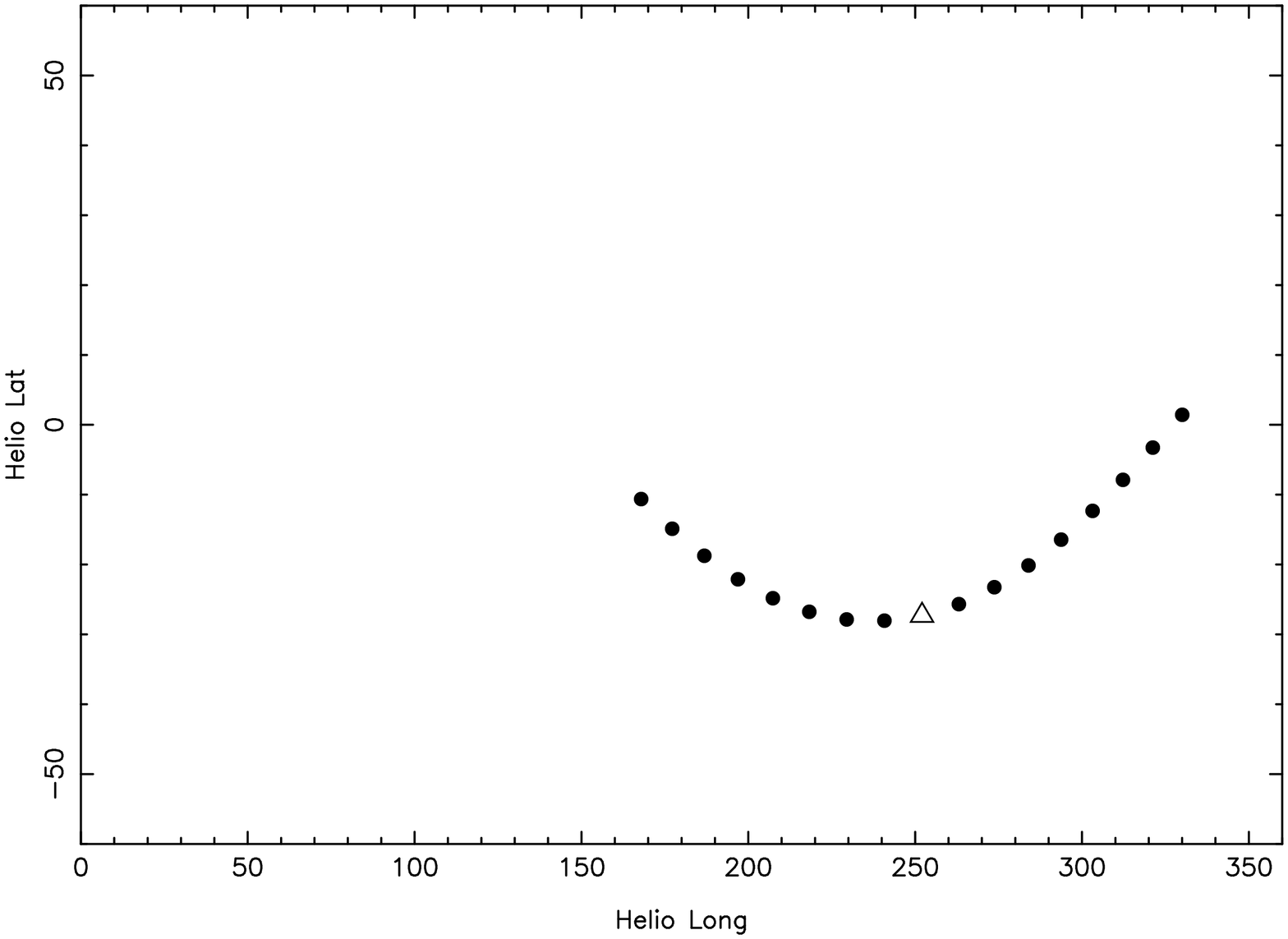}
\includegraphics[angle=-90,width=28pc]{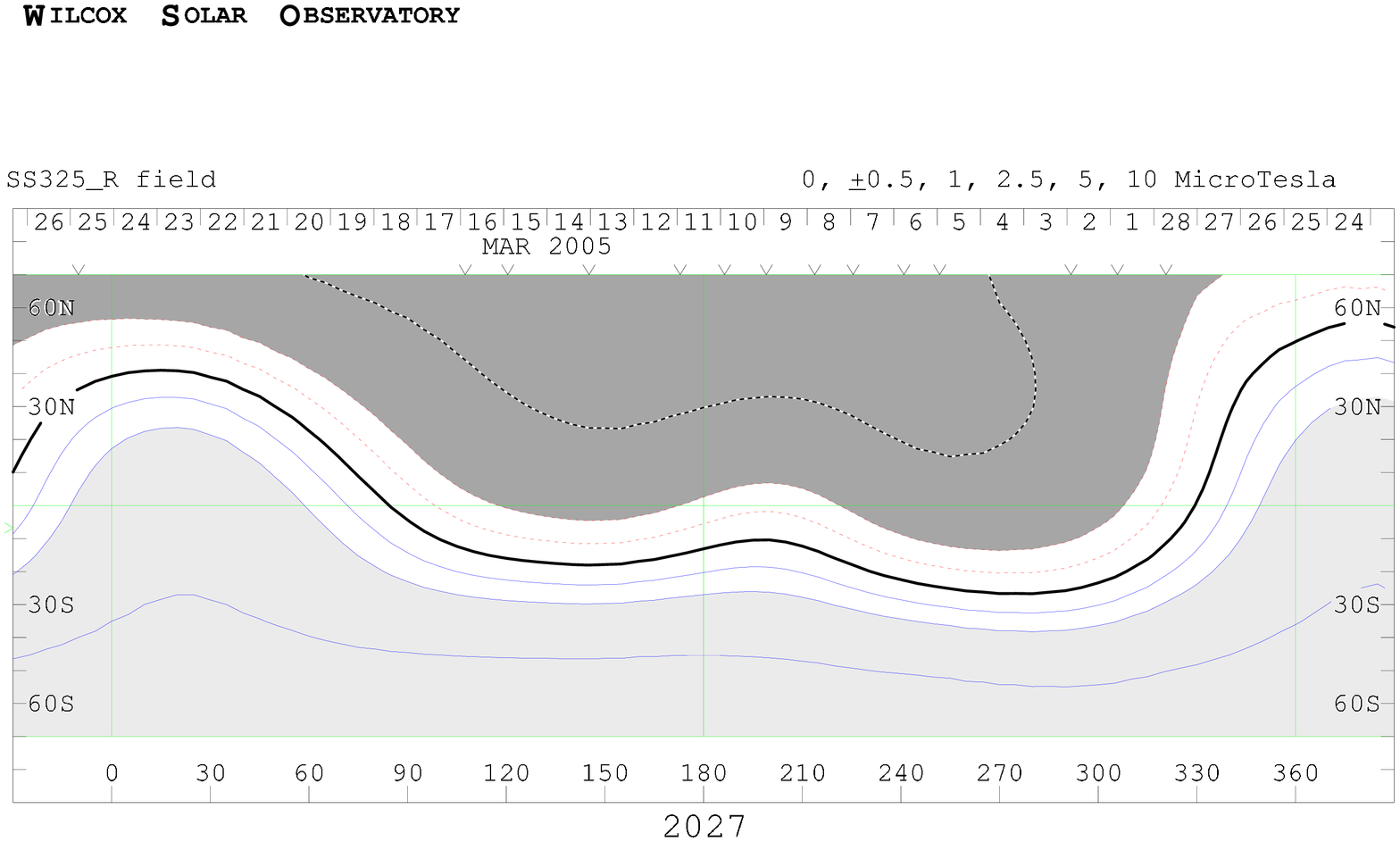}
\caption{Charts used in calculating the model rotation measures presented in Section 4.1.  (Upper panel) Projection of the line of sight of 2325-049 for March 12 on the surface of the Sun.  The triangular plotted point represents the point of closest approach to the Sun along the line of sight.  The solid points represent points at intervals of 10$^{\circ}$ in the angle $\beta$.  Each point also has a corresponding heliocentric distance. The progression along the line of sight from source to Earth is from right to left.  (Lower panel) Potential field model of the coronal magnetic field at $r = 3.25 R_{\odot}$ for Carrington Rotation 2027.  This chart was taken from the public data provided by the Wilcox Solar Observatory. }
\label{PFSS}
\end{figure}

\begin{figure}
\epsscale{0.70}
\includegraphics[angle=-90,width=35pc]{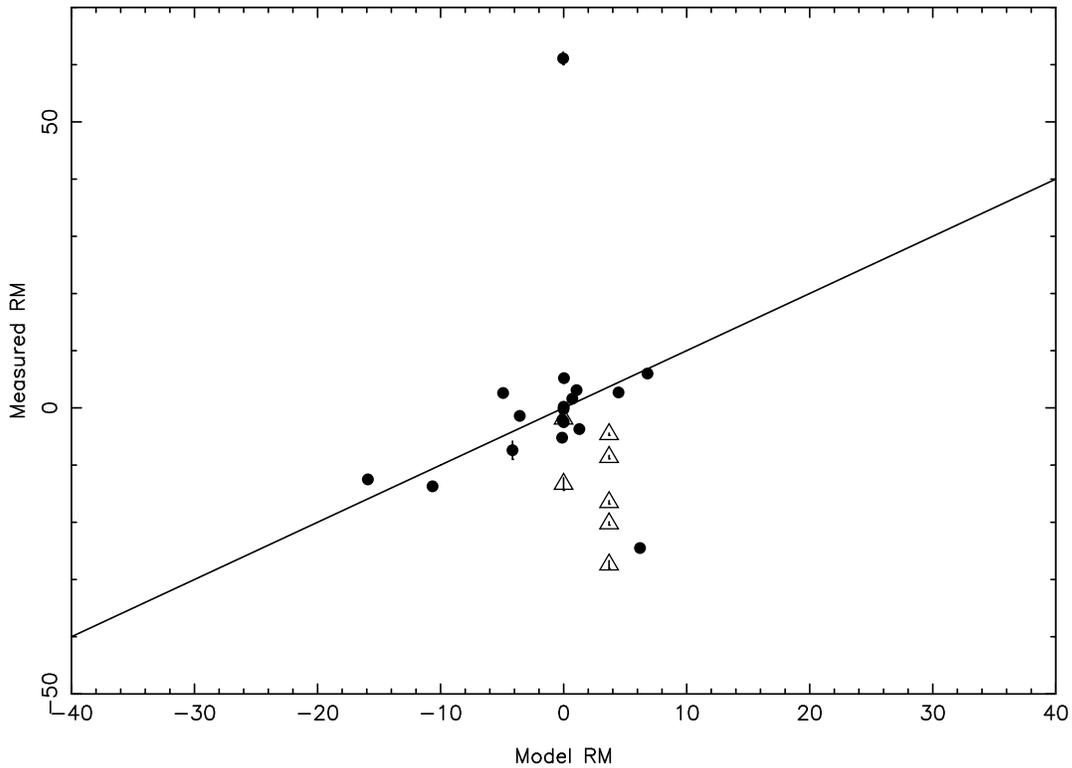}
\caption{Comparison of measured rotation measures with predictions described in Section 4.1 and expressed in eq (4).  Solid circles represent sources for which a session-mean rotation measure is sufficient to describe the coronal Faraday rotation.  The sets of triangles correspond to the two rotation measure-variable radio sources 2351-012 and 0046+067.  In these cases, the triangles represent the rotation measure during successive intervals within the 8 hour observing session.  The model rotation measures emergent from eq (6) have been adjusted by a systematic multiplicative scale factor of $\Gamma=0.475$ to improve agreement with observations.}
\label{SScormod}
\end{figure}

Our procedure in determining this first set of model rotation measures was as follows. 
\begin{enumerate}
\item For each source, we calculated the projection of the line of sight on the solar surface.  An example is shown in Figure~\ref{PFSS} for the case of the source 2325-049 on March 12.  The triangle symbol indicates the ``proximate point'', or point of closest approach of the line of sight to the Sun. 
\item The Wilcox Solar Observatory potential field estimates of the coronal magnetic field at a heliocentric distance of $r = 3.25 R_{\odot}$ were used to describe the coronal field at the time of the observations, available on the internet at (quake.stanford.edu/~WSO/WSO.html). The potential field model for Carrington Rotation 2027 is shown in Figure~\ref{PFSS}. For the analysis described in this subsection, the only information used for solar conditions at the time of the observations was the location of neutral line of the coronal magnetic field, taken from the potential field estimates. The charts were used to estimate the neutral line angle $\beta_c$.  
\item An estimate of the coronal Faraday rotation was made using a formula which can be obtained by substituting (2) and (3) into (1), and changing the variable of integration to the angle $\beta$ \citep{Patzold87}, 
\begin{equation}
RM = \left[  \frac{2CR_{\odot} N_0 B_0}{(\gamma -1) R_0^{\gamma-1}} \right] \cos^{\gamma -1}\beta_c
\end{equation}
where $\gamma \equiv \alpha + \delta$. The variable $R_0$ is the ``impact parameter'' in units of solar radii, and is defined in Figure~\ref{cartoon1}.  The sign of the rotation measure is determined by the polarity of the coronal field for $\beta < \beta_c$, and can be expressed by the sign of the variable $B_0$.  The constant $C \equiv \frac{e^3}{2 \pi m_e^2 c^4}$.
\end{enumerate}
A comparison of the predicted rotation measures from eq (6) with the measured rotation measures is shown in Figure~\ref{SScormod}.  The open triangles correspond to measurements for the rotation measure-variable sources 2351-012 and 0046+067.  Solid circles represent the remaining sources, for which a mean rotation measure for the entire session is sufficient.  The model calculations were done with values of $N_0 = 1.83 \times 10^6$ cm$^{-3}$, $B_0 = 1.01$ G, $\alpha = 2.36$ and $\delta = 2.0$, which appear reasonable on the basis of previous investigations \citep[e.g.][]{Spangler05}, but different plausible values would not change the nature of this figure. 

In addition, the model rotation measures plotted in  Figure~\ref{SScormod} were adjusted by a multiplicative scaling factor prior to be graphed.  A direct comparison of the observations with the model values emergent from eq (4) showed less agreement than shown in  Figure~\ref{SScormod} in the sense that the model values were predominantly larger than the measurements.  This  difference is most simply described by a systematic overestimate of the coronal density, the magnetic field strength, or both.  To compensate for such possible systematic overestimates, we have scaled {\em all} model rotation measures by a factor $\Gamma = 0.475$ before displaying them in  Figure~\ref{SScormod}.  A discussion of the significance of this value for the factor $\Gamma$ is discussed in Section 5.5 below.  

We also made a comparison between the observations and model values obtained from dual power law expressions for the density and magnetic field, as discussed in \cite{Mancuso00}. The expressions used were very slight modifications of expressions from \cite{Mancuso00}, specifically eq (6) of that paper for the density \citep[originally taken from][]{Gibson99} and (7) for the magnetic field. An expression for the model RM similar in form to (6) resulted (Spangler et al 2007, in preparation), and was compared with the data.  The resulting plot was essentially equivalent to Figure~\ref{SScormod}, provided that a slightly smaller scale factor $\Gamma=0.40$ is employed.

 Figure~\ref{SScormod} clearly shows substantial departures of the observed and model rotation measures in the cases of four lines of sight.  These are the two rotation-measure-variable sources, 2325-012 on March 19, and 0046+067 on April 1, as well as 2323-033 (data point near top of Figure), and 2352-016 (solid data point with most negative observed RM).  For the remaining 16 lines of sight, the observations show  fairly good adherence to the model line.  The root-mean-square (rms) departure of the 16 measured RMs from the corresponding model values was 3.35 rad/m$^2$ in a sample for which the range in RM was $\pm 20$ rad/m$^2$.  

The analysis described in this section therefore indicates that a simple coronal model contained in eq (2) and (3), together with the crucial ingredient of the location of the coronal neutral line along the line of sight (the parameter $\beta_c$), is sufficient to give a good estimate of the rotation measure in most cases.  An obvious addendum to this statement is that some type of coronal structure is responsible for producing large residuals in a sizable minority of observed cases. This has also been our experience in modeling previous coronal Faraday rotation projects.  
We did not undertake a comparison between our observations and more complex analytical models, specifically those which partition the line of sight into streamer and coronal hole portions, as was done in \cite{Mancuso00}.  Instead, in the next section we consider a model corona which incorporates much more empirical information about the state of the corona at the time of the observations.   
\subsection{Coronal Models with Data Input from Synoptic Maps}
The model presented in Section 4.1 assumes that $B$ and $n$ are functions only of heliocentric distance, with a change in the polarity of the coronal magnetic field at the neutral line as indicated by the potential field.  This is obviously a sweeping simplification of the true state of the coronal plasma.  The potential field estimate of the coronal field (see Figure~\ref{PFSS}) shows a gradual decrease of field magnitude as the neutral line is approached, rather than an abrupt reversal.  Of greater importance, the coronal density is not simply a function of heliocentric distance. As may be seen on coronagraph images, the density also depends on heliographic latitude and longitude, often showing an enhancement in the vicinity of the neutral line.  

Our second class of coronal models incorporated independent information on the heliographic latitude dependence of both the magnetic field strength and plasma density.  For coronal plasma density, we used estimates from the LASCO C2 coronagraph.  Data from the C2 coronagraph are available as data products\footnote{We thank Dr. Nathan Rich of the Naval Research Laboratory for advice and assistance with the use of these data.} in the form of  estimates of Carrington rotation-averaged plasma density at heliocentric distances of 3,4, and 5 $R_{\odot}$. The data giving the estimated, Carrington-rotation-averaged density or magnetic field strength at a prescribed heliocentric distance are referred to as synoptic charts of density or magnetic field, since they are conventionally displayed as charts or images.  We chose to use the density estimates from $r = 3.0 R_{\odot}$, which were closest in heliocentric distance to the potential field magnetic data described above.
The combination of the synoptic charts of coronal magnetic field at $r = 3.25 R_{\odot}$ and plasma density at $r = 3.00 R_{\odot}$ are our estimates of the factors in the integrand of eq (1).  

Model rotation measures using these synoptic charts were calculated in the following way.  It may be shown that equation (1) is equivalent to 
\begin{equation}
RM = C R_{\odot} R_0 \int_{LOS} n_e(\vec{r}) B(\vec{r}) \sec^2 \beta \sin \beta d \beta
\end{equation}
where $B$ is the magnitude of the assumed radial magnetic field, and $R_0$ is distance of closest approach of the line of sight in units of the solar radius ({\em not} the dimensional version of this quantity).  The angle $\beta$ is defined in Figure~\ref{cartoon1}.  

The estimates of plasma density and magnetic field must be extrapolated to the position of each point along the line of sight.  Our method of doing this is as follows.  Let $x \equiv r/R_{\odot}$.  We then define estimates of the density and magnetic field by 
\begin{eqnarray}
n_e(\vec{r}) = n_e(x,\beta) \equiv \Gamma_n n_3(\beta) (x/3.0)^{-2.36}  \\
B(\vec{r}) = B(x,\beta) \equiv \Gamma_b B_{325}(\beta) (x/3.25)^{-2.0}
\end{eqnarray}
The estimates given by equations (6) and (7) consist of extrapolating the density and magnetic field synoptic charts out into space radially, governed by the same power law dependence on the radial distance as assumed in eq (2) and (3).  Note that by writing $n_3(\beta)$ and $B_{325}(\beta)$, we mean that for a specific line of sight, there is a unique mapping between $\beta$ and the heliographic coordinates of the sub-solar point.  The factors $\Gamma_n$ and $\Gamma_b$ are scale factors which are employed to adjust the density and magnetic field estimates to agree with the observations. They obviously serve the same function as the parameter $\Gamma$ introduced in Section 4.1. By introducing these parameters, we are assuming that the synoptic charts correctly reproduce the variation of these plasma quantities with heliographic coordinates, but that the overall value of these quantities may not be accurate. In the present analysis, we will continue to distinguish $\Gamma_n$ and $\Gamma_b$, although a Faraday rotation measurement can only specify their product $\Gamma = \Gamma_n \Gamma_b$.   

The power law dependences on $x$ given in equations (8) and (9) have been previously used by \cite{Sakurai94a,Mancuso00} and \cite{Spangler05}.  Since $x=R_0 \sec \beta$, use of equations (6) and (7) converts (5) to an integration over the angle $\beta$.  The expression used to model the rotation measure is 
\begin{equation}
RM = \frac{CR_{\odot}(X_0)^2 \Gamma}{R_0} \int_{-\pi/2}^{+\pi/2}n_3(\beta)(x/3.0)^{-2.36}B_{325}(\beta)\sin \beta d \beta
\end{equation}
The parameter $X_0$ is the dimensionless heliocentric distance at which the potential field model is specified, which is $X_0=3.25$ in the present case. The explicit $x$ dependence of the density in (8) has not been completely removed, as it could have been.  The algorithm used to numerically evaluate (8) adopted the form shown to allow different functions (other than a single power law) for the $r$ dependence of $n_e(r)$.  

A set of model rotation measures was calculated for each line of sight, using a trajectory such as that shown in Figure~\ref{PFSS} for each line of sight, and the synoptic charts of coronal magnetic field  and plasma density.  To obtain overall agreement for the bulk of the data, a factor $\Gamma = 25$ was applied in equation (8) . Values of $\Gamma$ in the range $20 \leq \Gamma \leq 30$ yield approximately equivalent agreement between model and observed values.  This indicates that the corona in the region sampled by our observations, $5 \leq r \leq 10 R_{\odot}$ and beyond, is denser and has a stronger magnetic field than would be indicated by the independent magnetic field and density estimates for $r \simeq 3.0 R_{\odot}$, and the outward extrapolation according to equations (6) and (7).   

\begin{figure}
\includegraphics[angle=-90,width=28pc]{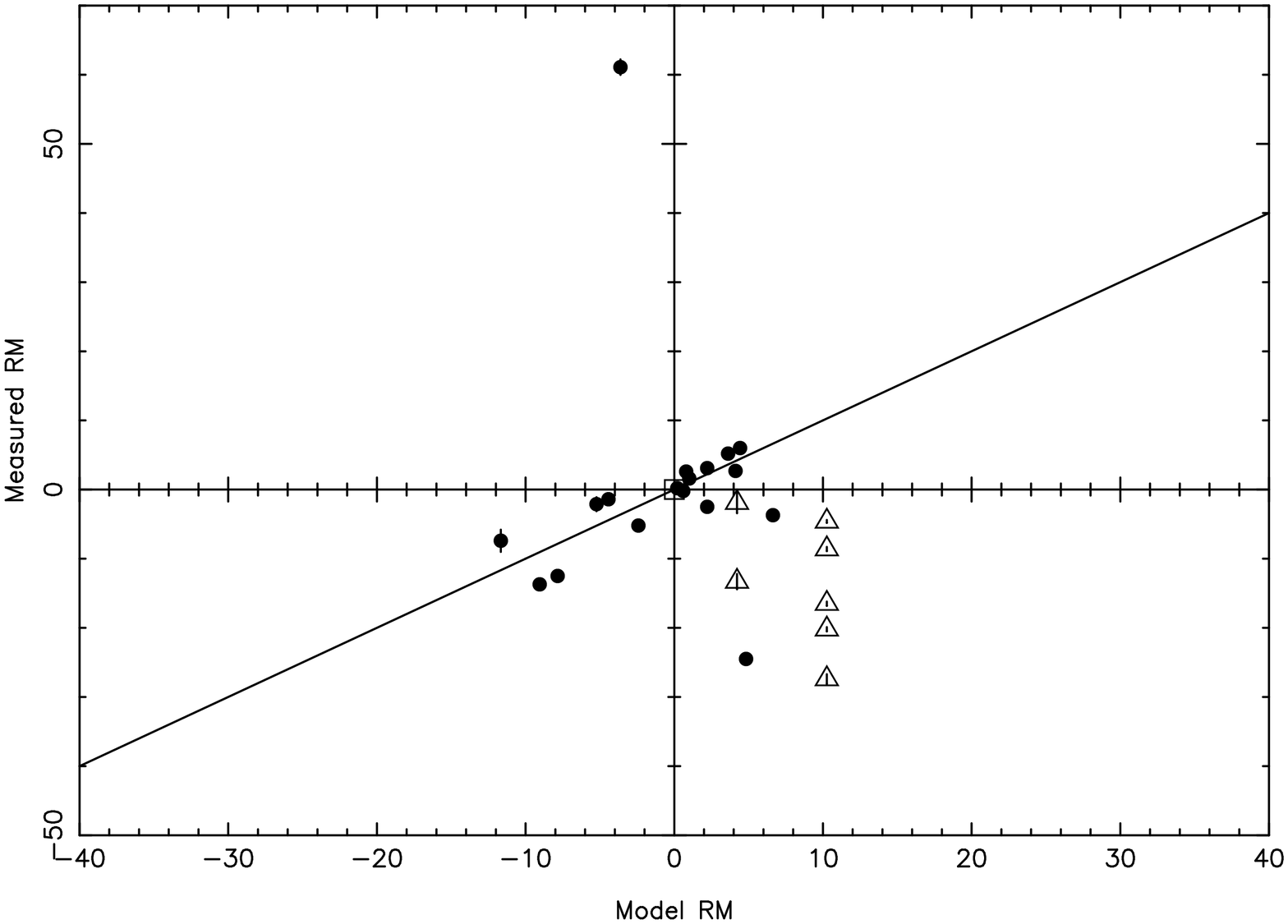}
\includegraphics[angle=-90,width=28pc]{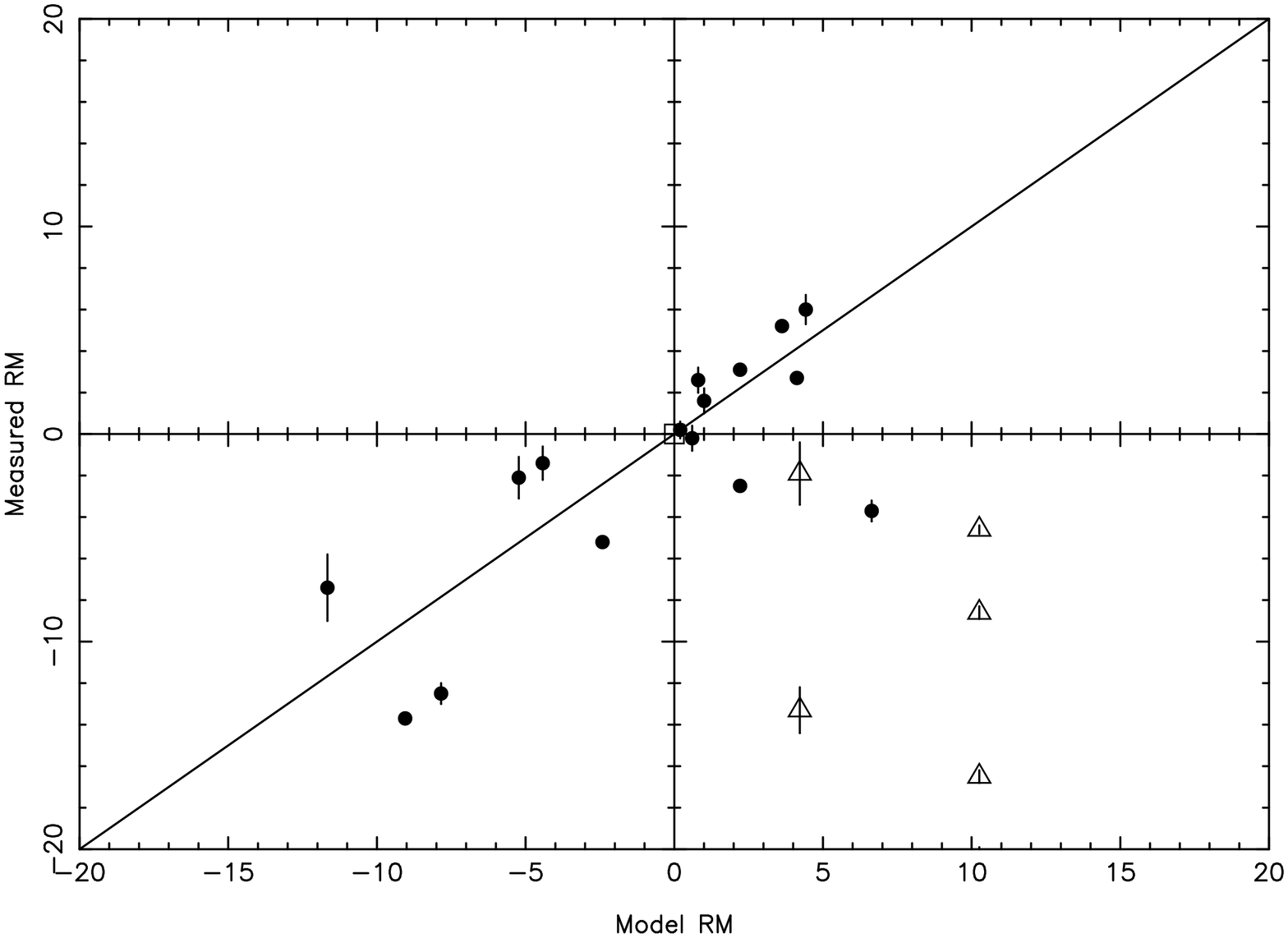}
\caption{Comparison of model and observed rotation measures.  The line through the origin is not a fit, but indicates perfect agreement.  The upper panel shows all the data.  The lower panel shows the inner part of the diagram, containing most of the measurements, for which $|RM| \leq 20$ rad/m$^2$. The sets of triangles indicate measurements of the two rotation measure-variable sources, 2351-012 and 0046+067.  In these cases, each plotted symbol indicates a measurement integrated over a time-limited subset of the data. The model rotation measures have been calculated with a value of the adjustment parameter $\Gamma = 25$ (see eq (8) for a definition of $\Gamma$).  }
\label{EMcormod}
\end{figure}

Figure~\ref{EMcormod} shows the comparison between the observed rotation measures and the model values, with the scale factor $\Gamma = 25$ applied. The physical significance of such a value for $\Gamma$ is discussed in Section 5 below.  The upper panel shows all the measurements, while the lower panel shows only those for which for which $|RM| \leq 20$ rad/m$^2$.  From Figure~\ref{EMcormod}, it may be seen that as was the case for the analysis of Section 4.1, four of the lines of sight have highly discrepant rotation measures (observed values far different from the modeled values). These are the same sources as noted in Section 4.1, 2323-033, 2352-016, and the two variable sources, 2351-012 and 0046+067). The data represented by triangles correspond to rotation measure-variables sources.  As before, these sources have several measured rotation measures for a single model value.   

Possible reasons for the four anomalous measurements are discussed in Section 4.4 below.  
For the remaining 16 lines of sight, there is a clear tendency for the data points to adhere to the line of agreement between the model and observed rotation measure. 

The degree of agreement between the model rotation measures and the data for the 16 selected lines of sight is comparable to that in the analysis of Section 4.1.  In fact, and somewhat surprisingly, the scatter of the meausrements about the solid line in Figure~\ref{EMcormod} is slightly larger than that in Figure~\ref{SScormod}, with an rms deviation of 3.71 rad/m$^2$ as opposed to 3.35 rad/m$^2$ for the analysis of Section 4.1.  While these values for the rms deviation are not significantly different, the more detailed analysis of this section has not provided an improved agreement between models and observations.  

\subsection{Comparison with Previous Results on RM Modeling}  There have now been three projects by our group in which VLA observations have been used to model the coronal magnetic field.  Mancuso and Spangler (2000) found that models similar to those used in Section 4.1 yielded a reasonable representation of the data, provided that the models incorporated a high density streamer belt centered on the neutral line as indicated from potential field maps.  Those authors found that for the best reproduction of the observations, it was necessary to have the width of the streamer belt vary with heliographic longitude, with a typical half width of $12\arcdeg$ \citep[See Figure 5,][]{Mancuso00}.  For the models of \cite{Mancuso00}, the most important input parameters from solar data was the position of the neutral line, and the width of the streamer belt. The agreement between model and observed rotation measures in \cite{Mancuso00} was similar to that in Figures~\ref{SScormod} and~\ref{EMcormod}, once the variable and highly discrepant lines of sight are removed from the latter figures.  

An analysis of two days of observations of the radio source 3C228 in August, 2003 (partially reported in \cite{Spangler05}; complete report in preparation) found that models of the sort presented in Section 4.1 reproduced the sign of the rotation measure and the sign of the temporal derivative of the rotation measure on two lines of sight which were continuously monitored for an eight hour observing session.  In addition, those models could reproduce the approximate magnitude of the RM for one of the lines of sight (that to the source 3C228 on August 16, 2003).  Unexpectedly, more sophisticated models which incorporated a finite thickness to the denser streamer belt did no better in reproducing the observations, and in fact did worse for one of the lines of sight (3C228 on August 18, 2003). 

Justin Kasper has analysed these  data by using MHD models of the corona to calculate the model rotation measures (Spangler et al (2007), in preparation). These MHD models incorporate contemporaneous photospheric magnetic field data.  The resultant model rotation measures are as successful as the analytic models in accounting for the observations, but not significantly more so.  One advantage of the Kasper model was greater success in matching the time derivative of, and total range in, the rotation measure time series for 3C228 on August 16, 2003 \citep[data presented in Figure 3 of][]{Spangler05}.  

The results of the present study are similar, in that use of more detailed coronal density and magnetic field information does not produce a major improvement on the algebraic model contained in eq (4).  Further discussion of this matter is given in Section 6.     

Before leaving this section,  it is worth noting that Mancuso (2006) recently has found that agreement between models and observations for a portion of the data set in Mancuso and Spangler (2000) is significantly improved if account is taken of the $7\arcdeg$ tilt of the magnetic axis of the Sun with respect to the rotation axis (which defines heliographic coordinates).  Mancuso (2006) found that when this angle was included in models of the sort discussed in Section 4.1, the agreement in  Figure 5 of Mancuso and Spangler (2000) was substantially improved. An analysis of this sort with the current data set will be undertaken in the near future.     

\subsection{Possible Explanations of Anomalous Rotation Measures}
The anomalous lines of sight are those for which the measured rotation measures differed drastically from the model predictions of Sections 4.1 and 4.2 (Figures~\ref{SScormod} and ~\ref{EMcormod}). 
We examined various data from the LASCO coronagraph to see if there were obvious, transient coronal features which might be present for these four sources and explain the large discrepancies or RM variability.  The most obvious possibility would be coronal mass ejections (CMEs).  We first searched the list of CME events maintained by the Naval Research Laboratory
\footnote{The internet URL used was lasco-www.nrl.navy.mil/cmelist.html.}. Examination of this list showed no evidence that a CME was responsible for any of the anomalies.  An easily visible loop CME occurred on March 12, but was on the opposite side of the Sun from the anomalous source 2323-033.  Also, the coronal rotation measure for 2323-033 was nearly constant over the eight hour observing period, whereas one would expect substantial temporal variations if the line of sight had been occulted by a CME.  

Two of the sources observed on March 19 were anomalous, 2352-016 with a roughly constant RM which departed significantly from the expected value, and 2351-012 with a variable RM.  These two sources had different types of anomalies, even though they are separated by only 17 degrees in position angle, and would seem to be probing the same part of the corona.  Examination of the C2 and C3 coronagraph data shows bright, very narrow streamers in the approximate directions of these two sources.  There also was a streamer near the line of sight for 3C228 on August 16, 2003, and a large and variable rotation measure was observed for that source \citep{Spangler05}.  

Thus a cursory examination of the independent data on the coronal state for the anomalous sources suggests a connection with coronal streamers which occult the line of sight.  \cite{Woo97} has long claimed a relationship between streamers and anomalies in radio scattering.  In this context, however, one must recall the results of \cite{Patzold98}, which showed that plausible models of streamers yield excess rotation measures which are small and slowly variable. 
  
\section{New Estimates of the Coronal Magnetic Field}
The most important observational result of this paper is contained in Figures~\ref{SScormod} and ~\ref{EMcormod}, which show reasonable agreement between measured and model rotation measures for a majority (16 of 20) of the lines of sight probed.  This agreement allows quantitative information to be extracted on the form of the coronal magnetic field.  In what follows, we concentrate on the conclusions to be drawn from the modeling in Section 4.2.  The case with the analytic models of Section 4.1 will be discussed in Section 5.4.  The  agreement in Figure~\ref{EMcormod} is dependent on a net adjustment factor $\Gamma = \Gamma_n \Gamma_b =25$. Overall scale factors in the range 20-30 give similarly good agreement.  This result in itself is important, because it indicates that the extrapolation  of estimated plasma parameters at $r \simeq 3.0 R_{\odot}$ into the outer corona are substantially lower than the observationally-required values.  Apparently the corona at distances of $5.0 R_{\odot} \leq r \leq 10.0 R_{\odot}$ is denser and/or has a stronger magnetic field than indicated by estimates from lower altitudes.  

In the remainder of this section, we consider what can be inferred about the coronal magnetic field.  Figure~\ref{EMcormod} indicates that the functional form of the field, that is being radial and having the dependence on heliographic latitude and longitude shown in Figure~\ref{PFSS}, is accurate or at least acceptable.  However, the magnitude of the field given by eq (7) with $\Gamma_b=1$ is too low.  To determine the magnetic field, we need to independently estimate the adjustment factor for the density, $\Gamma_n$.  This was done in three ways, which are summarized below.  

\subsection{Matching of Peak Densities in Synoptic Chart} Previous investigations in this program, as well as the present study, have used independent models for the plasma density as a function of heliocentric distance. A discussion of the models we have employed is given in Section 3 of \cite{Mancuso00}.  Mancuso and Spangler (2000) used an expression for the coronal streamer density, given in eq (6) of their paper,  due to Gibson et al (1999).  This expression would give a density of $2.36 \times 10^4$ cm$^{-3}$ at a characteristic distance of $6.2 R_{\odot}$.  The highest densities encountered on the synoptic chart for Carrington Rotation 2027, also extrapolated via eq (8) to $6.2 R_{\odot}$, are $\simeq 10.8 \times 10^4$ cm$^{-3}$.  The ratio of these two numbers may be taken as an estimate of $\Gamma_n$, which is then 2.18.  

A similar approach would be to take the densities from the synoptic chart and extrapolate them downward to the coronal base according to an $r^{-2.36}$ law.  This would give a peak base density of $8.0 \times 10^5$ cm$^{-3}$. This is low compared to the value of $1.83 \times 10^6$ cm$^{-3}$ which is used in eq (2) as well as our earlier investigations. Making the expressions equivalent would require a value of $\Gamma_n=2.29$ in eq (6), which serves as our second estimate of this quantity.  The two estimates for $\Gamma_n$  are obviously in completely agreement, and so we adopt a value of $\Gamma_n = 2.20$ as the result from matching analytic expressions for the streamer density as a function of heliocentric distance to the {\em peak} density from the synoptic chart.  

This value of $\Gamma_n$, together with a value of 25 for the net adjustment parameter $\Gamma$ yields our first estimate for the magnetic field adjustment factor $\Gamma_b = 11.4$.  In this case, the potential field values would underestimate the strength of the coronal field at greater heliocentric distances by more than a factor of 10.  The synoptic charts of the potential field for Carrington Rotations 2027 and 2028 give a maximum value of about $2.5 \times 10^{-2}$ G on the source surface at $r=3.25 R_{\odot}$.  Use of such a value for $B(r=3.25 R_{\odot})$ in eq (7), with an adjustment factor of $\Gamma_b = 11.4$  would give a value of $B = 7.8 \times 10^{-2}$ G for the coronal field at a fiducial distance of $r=6.2 R_{\odot}$.  This is exactly twice the value of the field at this heliocentric distance reported by \cite{Spangler05}, based on other VLA observations made in 2003,  but is only slightly greater than the value of $B(r=6.2 R_{\odot}) = 6.2 \times 10^{-2}$ G which would result from the model expression given in \citep[eq(7)]{Mancuso00}.  

\subsection{Comparison of Columnar Depths: Method I} 
The method for estimating $\Gamma_n$ (and thus 
$\Gamma_b$) in Section 5.1 relied on matching the peak density on the synoptic chart with model analytic expressions for the plasma density as a function of heliocentric distance.  However, those model expressions were generally obtained from measurements of a path integral of the density along the line of sight, such as the arrival time delay as a function of frequency from a trans-solar radio beacon \citep{Bird94,Bird96}.  This parameter, referred to as the columnar depth by Bird and colleagues, is also termed the dispersion measure (DM) in applications to pulsar signals which have propagated through the interstellar medium. Even in the case in which separate determinations were made of the dispersion measure in the streamer and coronal hole plasmas, the retrieved density represented an average over regions in which the density was more or less than the mean value.  Our second method of determining $\Gamma_n$ therefore consisted of comparing the DM (dispersion measure) values corresponding to the streamer densities reported by  \citep{Bird94,Bird96} with those directly calculated from the density synoptic chart.  

First we consider the expression for the dispersion measure corresponding to an algebraic density law of the form given in eq (2).  The dispersion measure is given by
\begin{equation}
DM = \int_{L} n_e(\vec{r}) dz
\end{equation}
As in the case of rotation measure, this can be transformed to an integral over the angle $\beta$, 
\begin{equation}
DM = (R_{\odot} R_0) \int_{-\pi/2}^{\pi/2} n_e(\vec{r}) \sec^2\beta d \beta
\end{equation}
where $R_0$ is the point of closest approach in units of the solar radius, {\em not} its dimensional counterpart.  

We now adopt an expression for $ n_e(\vec{r})$ of the form given in eq (2).  The heliocentric distance can also be changed to a function of $\beta$, 
\begin{equation}
r = R_{\odot} R_0 \sec \beta
\end{equation}
We then substitute eq (11) into eq (2), and that into eq (10) above.  For simplicity of the algebra, we adopt $\alpha = 2.5$, which is quite close to the more precise value of $\alpha=2.36$ used throughout this paper.  We then have 
\begin{eqnarray}
DM = \frac{2 N_0 R_{\odot}}{(R_0)^{3/2}} \int^{\pi/2}_0 \frac{d \beta}{\sqrt (\sec \beta )} \\
DM = \frac{2 N_0 R_{\odot}}{(R_0)^{3/2}} \frac{(2 \pi)^{3/2}}{(\Gamma(\frac{1}{4}))^2}
\end{eqnarray} 
where $\Gamma(\frac{1}{4})$ in eq (13) is the gamma function, not the adjustment coefficient defined in Sections 4.1 and 4.2.  

Note that this expression has the dispersion measure $\propto R_0^{3/2}$. In what follows, we have adopted a value of $N_0 = 1.83 \times 10^6$ cm$^{-3}$ as the sole input parameter to this expression. This value has been used earlier in this paper, as well as repeatedly in other papers in this series. We compared the values for DM obtained from eq (13) with those from the model for the corona discussed in Section 4.2, and incorporating the density synoptic chart.  To do this, the density expression given by eq (6) (with $\Gamma_n=1$) was integrated along each of the 20 lines of sight, resulting in 20 estimates of the dispersion measure which described the synoptic coronal model of Section 4.2.  In each case, the estimate obtained from eq (13) was larger than that obtained by direct integration of the synoptic model through the corona.  To obtain our second estimate of the parameter $\Gamma_n$, the following procedure was done.  For each line of sight the ratio $R$ of the DM given by eq (13) to that obtained by direct, line of sight integration of the synoptic density model was calculated.  The median value of this ratio was 5.62, which we adopt as our second estimate of $\Gamma_n$. A plot (not shown) of the numerically-integrated DM values as a function of $R_{\odot}$ shows good adherence to a $R_0^{-3/2}$ relationship, and a majority (16/20) of values showed good agreement with eq (13) for the input parameter $N_0 = 1.83 \times 10^6$ cm$^{-1/3}$, provided that the empirical DM values were scaled up by a factor of $\Gamma_n=5.62$.  The remaining 4 points were consistent with an $R_0^{-3/2}$ functional form, the same value of $N_0$, and $\Gamma_n=3.1$.  These data points may represent lines of sight which passed through particularly dense parts of the corona.  

This value of $\Gamma_n=5.6$ is more than a factor of 2 larger than that in Section 5.1, and would imply a lower magnetic field adjustment factor $\Gamma_b=4.4$.  The corresponding fiducial value of the coronal magnetic field at $r=6.2 R_{\odot}$ is 30 mG.  This compares favorably with the estimate of 39 mG from observations of the corona 3 years earlier \citep{Spangler05}.   

\subsection{Comparison of Columnar Depths: Method II} The method described in Section 5.2 compared the  dispersion measures calculated from the extrapolated coronagraph data with those obtained from an analytic expression for the coronal density as a function of heliocentric distance.  Although this analytic expression (eq 13) was obtained from a body of dispersion measure data, a more direct approach would be to directly compare the synoptic model values obtained as described in Section 5.2 with dispersion measure measurements.  For this purpose, we used the results of \cite{Bird96}, which are ideally suited for this purpose.  The observations were made at superior conjunction of the Ulysses spacecraft in 1995.  As is the case for the present observations, the Sun was near the time of solar minimum (end of cycle 22 rather than end of cycle 23), and a large number of lines of sight were probed  (42 measurements of DM given in Table 1 of \cite{Bird96} were utilized), Some of these lines of sight passed through streamers, while others were mainly contained within coronal holes. To obtain our third estimate of $\Gamma_n$, the following procedure was used. 
\begin{enumerate}
\item The expression in eq (13) was divided by the median value of $R=5.62$  to provide a smooth representation of the dependence of the synoptic model dispersion measures on heliocentric distance. The corresponding value expected value of DM at $r=6.0 R_{\odot}$ was 3700 Hexems \footnote{A Hexem is defined as $10^{16}$ m$^{-2}$ = $10^{12}$ cm$^{-2}$.}. It is to be emphasized that eq (13) was solely used as a smooth fitting function to represent the numerically integrated DM estimates made from the synoptic charts. 
\item The 42 DM measurements listed in Table 1 of \cite{Bird96} were corrected to the value which would have been observed at 6.0 $R_{\odot}$, by scaling by the 3/2 power of the heliocentric distance.  For each of these measurements, the ratio to the reference value of 3700 Hexems was taken.  The mean of these ratios was 5.25 and the median was 4.97.  We adopt the median value of this ratio as our third estimate $\Gamma_n=5.0$.  It is obviously in good agreement with the value emergent from the technique in Section 5.2. The corresponding value of $\Gamma_b$ is 5.0, with an associated magnetic field strength at $6.2 R_{\odot}$, calculated as in Section 5.2, of 34 mG. 
\end{enumerate} 
\subsection{Summary of Results for Magnitude of Coronal Magnetic Field} 
We thus have three estimates of the parameter $\Gamma_n$, with corresponding values of the parameter which is of greater interest to us, $\Gamma_b$.  The results of this section, together with the results from previous sections, thus indicate that the magnetic field in the corona at distances of $r \geq 5 R_{\odot}$ is between 4.4 (Section 5.2) and 11.4 (Section 5.1) times the value that would be obtained by an extrapolation of the potential field  out into the corona according to an inverse square law.  The corresponding range in values of the coronal magnetic field at the fiducial distance of $6.2 R_{\odot}$ is 30 - 78 mG.  This range is in good agreement with the recent discussion by Spangler (2005). 

The choice of a fiducial heliocentric distance of $6.2 R_{\odot}$ was made to permit ready comparison to the results in \cite{Spangler05}, where it was an important characteristic of the observations presented.  To report our results in a more convenient form, we scale the magnetic field estimates to a heliocentric distance of $5 R_{\odot}$ according to an $r^{-2}$ dependence.  The total range in estimates for $B(r=5 R_{\odot})$ is then 46-120 mG, with a more restricted range of 46-52 mG resulting from the analyses of Sections 5.2 and 5.3.  

Within the total range of deduced parameters, we prefer the lower values ($\Gamma_b \simeq 4.4-5.0$ with $B(r=6.2 R_{\odot}) = 30 - 34$ mG; $B(r=5.0 R_{\odot}) = 46 - 52$ mG ) for the following reasons.  First, as mentioned above, the coronal density profiles used in those calculations  were obtained from estimates and measurements of the dispersion measure, and therefore represent an average over high and low densities along the lines of sight. Accordingly, methods of determining $\Gamma_n$ which scale the coronal model densities to the dispersion measure (methods described in Sections 5.2 and 5.3) appear preferable.  Second, MHD simulations of the solar wind which incorporate coronal data as a boundary condition require coronal fields about a factor of 4 higher than the potential field values in order to match solar wind conditions at 1 a.u., particularly the value of the interplanetary magnetic field (Ofer Cohen, University of Michigan, private communication).  The methods in Sections 5.2 and 5.3 would be consistent with this conclusion.  

The discussion in this section demonstrates that we are converging on a good knowledge of the magnetic field in the solar corona.  However, it also illustrates that future analyses of this sort, either with new Faraday rotation observations or with reanalysis of existing data, would  greatly benefit from the best possible estimates of the coronal plasma density.  

\subsection{Compatibility of Different Rotation Measure Modeling Schemes}
The final, remaining issue in this section is the reconciliation of the models of Sections 4.1 (simple analytic models for the coronal plasma structure) and Section 4.2 (use of empirical synoptic charts of coronal plasma properties).  The analysis of Section 5 resulted in corrections to the independent plasma density and magnetic field estimates which moved them towards those adopted in equations (2) and (3) of Section 4.1.  Nonetheless, eq (4), obtained from equations (2) and (3), tended to {\em overestimate} the rotation measure for most lines of sight, and was adjusted downwards by $\Gamma \simeq 0.5$ to achieve agreement with the bulk of the observations.  Given the results earlier in this section, we would have expected $\Gamma$ for the simple analytic models to have been closer to unity.  

It is possible to be cavalier and maintain that a factor of 2 in $\Gamma$ is acceptable, given the incompletely-specified, real state of the corona and the simplifications implicit in eq (4).  Furthermore, the analysis of Section 4.2 incorporated much more detailed information on the corona, albeit with applied scale factors as discussed previously in this section, and therefore can be expected to yield more accurate results.  Nonetheless, a plausible explanation for the apparent preferred value of $\Gamma \simeq 0.5$ for the analysis of Section 4.1 is as follows.  
The derivation leading to eq (4) assumes that the coronal field discontinuously changes from $B(r)\hat{e}_r$ to  $-B(r)\hat{e}_r$ (or vice-versa) at the coronal neutral line.  In contrast, as may be seen in Figure~\ref{PFSS}, the potential field model has a smooth decrease in the field magnitude as the neutral line is approached. For much of a typical line of sight, the magnetic field will be less than an outward, $r^{-2}$ extrapolation of the maximum field strength seen in Figure~\ref{PFSS}. Since the plasma in the vicinity of the neutral line can dominate the contribution to the total rotation measure, an overestimate of the field magnitude in the vicinity of the neutral line can produce an overestimate of the model rotation measure.  Although we do not pursue this further in the present paper, the previous considerations indicate that the factor $\Gamma$, as applied to the analysis of Section 4.1, might contain  information on the form of the coronal field near the neutral line, and the extent to which it departs from a potential field.  

\section{Summary and Conclusions}
The conclusions of this paper are as follows. 
\begin{enumerate}
\item Dual frequency, 1.465 and 1.665 GHz polarimetric measurements were made with the VLA of 20 sources whose lines of sight passed at heliocentric distances of 5.6 to 9.7 $R_{\odot}$. The observations, made in March and April 2005 (Carrington Rotations 2027 and 2028) yielded measurements of the Faraday rotation measure (RM) along these 20 lines of sight.  
\item The mean rotation measures observed ranged from -24.5 to 61.1 radians/m$^2$, with most having $|RM| \leq 10$ rad/m$^2$.  
\item The observed rotation measures were compared with a simple analytic model for the coronal plasma, with a radial magnetic field, and power law dependences of the density and magnetic field strength on heliocentric distance.  The only input from observations of the corona on the days of observation was the location of the coronal magnetic neutral line.  A formula (eq (4)) resulting from this analytic model did a fair job of reproducing the observed RMs for most (16 of 20) of the lines of sight, although a multiplicative scale factor of $\Gamma=0.48$ had to be applied to the model RMs.   
\item The measurements were also compared with a model which made more extensive use of independent data on coronal conditions, specifically the potential field model of the magnetic field defined on a source surface at $r = 3.25 R_{\odot}$, and density estimates from the C2 coronagraph data, defined at $r = 3.00 R_{\odot}$. For calculations of magnetic field and density throughout the corona, these values were extrapolated according to power laws of the heliocentric distance.  
\item Sixteen of the 20 lines of sight were found to be in reasonable agreement with the predictions of the aforementioned model, when allowance was made for an overall scaling of the model densities and magnetic field.  The net adjustment factor is $\Gamma=25$, with equally acceptable factors $20 \leq \Gamma \leq 30$.  
\item Four of the 20 lines of sight were completely discordant with this model. Two of these  lines of sight showed major Faraday rotation variability during the session.  The anomalous lines of sight do not appear to be associated with coronal mass ejections, but there may be an association with coronal streamers.  
\item Much of the analysis effort consisted of extracting the density and magnetic field scale factors $\Gamma_n$ and $\Gamma_b$, which determine the overall RM scale factor $\Gamma=\Gamma_n \Gamma_b$. Three methods were employed to estimate the adjustment factor for the plasma density $\Gamma_n$, which is the factor by which the density model described in point \# 4 must be multiplied to be in agreement with the observations.  An estimate of $\Gamma_n$ is necessary to extract the corresponding value  $\Gamma_b$ for the magnetic field. The values for $\Gamma_n$ ranged from 2.2 to 5.6, with corresponding values of $\Gamma_b$ of 4.4 - 11.4.  The resulting value for the coronal magnetic field at a fiducial distance of $6.2 R_{\odot}$, outside of the region around the neutral line, ranges from 30 to 78 mG. At a more convenient reference distance of $5.0 R_{\odot}$ the corresponding range in the magnitude of the magnetic field is 46-120 mG.  
\item For reasons discussed in Section 5, we favor the larger values for $\Gamma_n$, and lower values for $\Gamma_b$ and the magnetic field strength in the corona.  The result of the present paper is therefore a value of the coronal magnetic field (again, outside of the region around the neutral line) of 30 - 34 mG at a fiducial heliocentric distance of $6.2 R_{\odot}$.  This is in reasonable agreement with the value of 39 mG obtained by \cite{Spangler05} on the basis of independent VLA observations made in 2003. At a heliocentric distance of $r = 5.0 R_{\odot}$, this preferred range is 46 - 52 mG.  
\end{enumerate}

\acknowledgments
This work was supported at the University of Iowa by grants ATM03-11825, and ATM03-54782 from the National Science Foundation. Laura Ingleby was also supported by a James Van Allen Undergraduate Research Grant from the Department of Physics and Astronomy, University of Iowa. The authors appreciate Professor Van Allen's interest in and encouragement of this project.  Catherine Whiting was also supported by the University of Iowa through the Undergraduate Scholar Assistant program. SRS acknowledges very useful discussions with Ofer Cohen of the University of Michigan about the strength of the coronal magnetic field during the 2006 NSF/SHINE Workshop in Midway, Utah.  Finally, we thank Nathan Rich and Angelos Vourlidas of the Naval Research Laboratory for their help in using the SOHO/LASCO C2 data.

\end{document}